\documentclass[preprintnumbers,showpacs,amsmath,amssymb,floatfix,prd,superscriptaddress,nofootinbib]{revtex4}
\usepackage{graphicx}
\usepackage{epsfig}
\usepackage{bm}
\usepackage{amsfonts}

\begin{document}

\title{A vacuum component of the Universe must evolve}

\author{Vladimir Burdyuzha }
\email{burdyuzh@asc.rssi.ru}\affiliation{Center for Advanced
Mathematics and Physics, National University of Sciences and
Technology,H-12,Islamabad, Pakistan} \affiliation{Astro-Space
Center, Lebedev Physical Institute, Russian Academy of Sciences,
Profsoyuznaya 84/32, 117997 Moscow, Russia }

\begin{abstract}
The evolution of the vacuum component of the Universe is
investigated in the quantum as well as the classical regimes.
Probably our Universe has arisen as a vacuum fluctuation and very
probably  it had a high symmetry for Planckian parameters.  In the
early epochs of its cooling, during the first second, the vacuum
component of the Universe had been losing its high symmetry by phase
transitions, since condensates of quantum fields carried negative
contributions (78 orders) to its positive energy density. After the
last phase transition (quark-hadron) the vacuum energy `has
hardened'. At this moment ($10 ^{-6}$ sec) its energy density can be
calculated using  Zeldovich's formula, inserting an average value of
the pseudo-Goldstone boson masses (pi-mesons) that characterizes
this chromodynamical vacuum.  The chiral symmetry was then lost. The
dynamics of the equilibrium vacuum after its `hardness' is
considered by applying the holographic principle. During of next
$4\times10^{17}$ sec the vacuum component of the Universe had been
losing 45 orders by the creation of new quantum states. Utilizing
the holographic principle, we solve the cosmological constant
problem because 123 crisis orders disappear in usual physical
processes. The density of vacuum energy from redshift  $z=0$ up to
redshift $z=10^{11}$ is also calculated in the classical regime of
the Universe evolution using  the ``cosmological calculator''.
\end{abstract}


\maketitle

\section{Introduction}

In this article we propose that a $\Lambda$-term, vacuum energy,
cosmological constant and dark energy have a common origin. The
research of the vacuum energy evolution presented a large interest
always. A. Einstein had introduced the $\Lambda$-term in his field
equations as a property of space-time \cite{1}
\begin{equation}\label{1}
G_{\mu\nu}+\Lambda g_{\mu\nu}=-8\pi G_NT_{\mu\nu}.
\end{equation}
If one puts the $\Lambda$-term in the right side of the above
equation (\ref{1}) then it will be a form of energy called dark
energy, due to the absence of a good explanation of its nature,
\begin{equation}\label{2}
G_{\mu\nu}=-8\pi G_NT_{\mu\nu}-\Lambda g_{\mu\nu}.
\end{equation}
The present value of this form of dark energy is:
\begin{equation}\label{3}
\rho_{DE}=\rho_\Lambda\sim10^{-47}(\text{GeV})^4\sim0.7\times10^{-29}\text{g/cm}^3,
\end{equation}
if $H = 70.5$ ($\text{kmsec}^{-1}$/Mpc). Furthermore, this form of
dark energy provides the reason of the present accelerated expansion
of our Universe (generally speaking our Universe is one of many
universes in a multiverse).  It is suggested that in the Planckian
epoch of the Universe evolution, this form of dark energy had the
density (UV cutoff):
\begin{equation}\label{4a}
\rho_\Lambda\sim2\times10^{76}(\text{GeV})^4\sim0.5\times10^{94}\text{g/cm}^3,
\end{equation}
for $M_{Pl}=1.2\times10^{19}$ GeV. From equations (\ref{3}) and
(\ref{4a}), a question arises why this huge difference takes place
in the present value of $\Lambda$-term to its value in the Planckian
epoch (123 orders)? This inexplicable difference caused a crisis of
theoretical astrophysics as mentioned in all reviews
\cite{2,3,4,5,6,7,8,9}, although many interesting hypotheses were
constructed to overcome this crisis
\cite{10,11,12,13,14,15,16,17,18,19,20,21,22,23,24,25}.

Probably, the most adequate explanation of dynamical (relaxation)
mechanism for $\Lambda$-term has been suggested by V. Rubakov
\cite{18}. Namely, the theory of primordial nucleosynthesis requires
that a large part of the vacuum energy had already reduced to
nucleosynthesis epoch. Therefore, the relaxation of the vacuum
energy should have occurred at some earlier cosmological stage.
Besides, the theory of formation of baryon structures in the
Universe requires a long matter dominated epoch that points also in
the same direction. The last observations showed that a parameter $w
= p/\rho$ characterizing dark energy is close to $-1$ with   $-0.14
< 1+w < 0.12$ \cite{26}.  But, in the early epochs during phase
transitions the $\Lambda$-term was not the cosmological constant. It
had become practically the cosmological constant only after the last
(quark-hadron) phase transition when the temperature of the Universe
dropped from $10^{19}$ GeV to 150 MeV. Before this in a positive
vacuum energy, condensates of quantum fields had carried negative
contributions as has already been noted (for existence of the large
scale baryon structure the small positive vacuum energy is only
possible \cite{27}). A. Dolgov was the first one who has discussed
this compensation hypothesis for the vacuum energy of the Universe
\cite{28}.

Probably, other time it is necessary to give the definition of
vacuum. A vacuum is defined as the stable state of quantum fields
without excitation of wave modes. In geometrical physics a vacuum is
the state in which the geometry of space-time does not deform. In
quantum cosmology a vacuum is condensates of quantum fields
appearing as the result of relativistic phase transitions. In
classical physics a vacuum is a world without particles and this
world is flat. The equation of state for a vacuum is $p = -\rho $.
At the first we consider a quantum regime of the vacuum evolution
and after we consider a classical one. A novel concept presented in
this article is that the Universe is expanding by losing vacuum
energy by the creation of new quantum states (45 orders during
$13.76\times10^9$ years). 45 orders of the classical regime and 78
orders of the quantum regime decreased vacuum energy of the Universe
on 123 orders and probably this is the solution of the cosmological
constant problem, if the cosmological constant is the vacuum energy.

\section{Phase transitions}
First of all we note that microscopic defects of a gravitational
vacuum which were produced in the quantum regime of the Universe
evolution contributed to the total energy of vacuum:
\begin{equation}\label{4}
\Lambda=\Lambda_\text{QF}+\Lambda_\text{GVC},
\end{equation}
here: QF are quantum fields, GVC is a gravitational vacuum
condensate \cite{27}. These microscopic topological defects
(worm-holes, micromembranes, microstrings, monopoles) had different
dimensions and might be a carrier of dark energy in very early
epochs also. Besides, the gravitational vacuum condensate fixed the
origin of time in our Universe \cite{29}. Unfortunately, we do not
know how exactly our Universe has been losing its high symmetry. The
elementary chain of the phase transitions, from which only the two
last ones can be calculated exactly, was described in our article
\cite{30}:
\begin{widetext}
\begin{eqnarray}\label{5}
&&P\Rightarrow D_4\times[SU(5)]_\text{SUSY}\Rightarrow
D_4\times[U(1)\times SU(2)\times SU(3)]_\text{SUSY}\nonumber\\&&
10^{19} \text{GeV} \ \ \ \ \ \ \ \ \ \ \ \ \ \ \ \  10^{16}
\text{GeV}
\\&& \Rightarrow D_4\times U(1)\times SU(2)\times SU(3)\Rightarrow
D_4\times U(1)\times SU(3)\Rightarrow D_4\times U(1)\nonumber\\&&
10^5-10^{10} \text{GeV}\ \ \ \ \ \ \ \ \ \ \ \ \ \ \ \ \ \ 100
\text{GeV} \ \ \ \ \ \ \ \ \ \ \ \ \ \ \ \ \ \ \ \ \ \ \  0.15
\text{GeV}\nonumber
\end{eqnarray}
\end{widetext}
The two last condensates of quantum fields in the framework  of
Standard Model ($\Lambda_\text{SM}$)  may be calculated. They have
an asymptotic equation of state $p= -\rho$ and they are named the
Higgs condensate in the theory of electro-weak interaction
($\rho_\text{EW}$) and the quark-gluon condensate in quantum
chromodynamics ($\rho_\text{QCD}$). Therefore:
\begin{equation}\label{6}
\Lambda_\text{QF}=\Lambda_\text{EW}+\Lambda_\text{QCD},\ \
\Lambda_\text{QF}=-\rho_\text{EW}-\rho_\text{QCD}
\end{equation}
In our article \cite{27} we have already given a value of
$\rho_\text{EW}$ as $\Lambda_\text{SM}$:
\begin{equation}\label{7}
\rho_\text{EW}=-\frac{m_H^2m_W^2}{2g^2}-\frac{1}{128\pi^2}(m_H^4+3m_Z^4+6m_W^4-12m_t^4).
\end{equation}
For mass of Higgs $m_H  \sim 160$ GeV we have:
\begin{equation}\label{8}
\rho_\text{EW}\sim -(120 \text{GeV})^4.
\end{equation}
This estimate was obtained in the article \cite{9}. But, the most
interesting condensate for us is the quark-gluon one since at this
moment of the Universe evolution the vacuum energy `has hardened'.
In article \cite{9}, the estimate of energy density of the
quark-gluon condensate is also presented:
\begin{equation}\label{9}
\rho_\text{QCD}\sim -(265 \text{GeV})^4.
\end{equation}
Note that only the quark-hadron phase transition quenches more than
10 orders of the 78 orders.
\begin{equation}\label{10}
\Big(\frac{120}{0.265}\Big)^ 4 \sim 4\times10^{10},\ \
\Big(\frac{M_{ Pl}}{ M_{ QCD}}\Big)^ 4 \sim4.5 \times10^{ 78}
\end{equation}
Unfortunately, the remaining contributions in the beginning and in
the middle of the chain of relativistic phase transitions (\ref{5})
are not possible to calculate exactly. Besides, the initial stage
might be more complicated. For example: $P\Rightarrow  E_6
\Rightarrow    O(10)\Rightarrow   SU(5)\ldots$. Whereas the last
chromodynamical phase transition (QCD) was investigated in the
review \cite{31} extensively. The chiral QCD symmetry $SU(3)_ L
\times SU (3)_ R$ is not an exact one and pseudo-Goldstone bosons
are the physical realization of this symmetry breaking. The
spontaneous breaking of the chiral symmetry leads to the appearance
of an octet of pseudoscalar Goldstone states in the spectrum of
particles. For the temperature of the chiral symmetry breaking ($T_
c\sim$ 150 MeV) the main contribution in the periodic collective
motion of a nonperturbative vacuum condensate determined pi-mesons
as the lightest particles of this octet. In this process pi-mesons
are excitations of the ground state and they definitely characterize
this ground state (that is they characterize the QCD vacuum). And
density of this vacuum energy may then be calculated.

Ya. Zel'dovich\cite{32} attempted to calculate a nonzero vacuum
energy of our Universe in terms of quantum fluctuations of fields as
a high order effect 40 years ago. He inserted the mass of proton or
electron in his formula but the result was not satisfactory. The
situation has changed since then if the average mass of pi-mesons
($m_\pi$  = 138.04 MeV) is inserted and  N. Kardashev's modification
\cite{33} of Ya. Zeldovich's expression is used:
\begin{equation}\label{11}
\Lambda=8\pi G_N^2m_\pi^6h^{-4}\ \text{cm}^{-2},
\end{equation}
\begin{equation}\label{12}
\rho_\Lambda=G_Nm_\pi^6c^2h^{-4}\ \text{gcm}^{-3},
\end{equation}
then
\begin{equation}\label{13}
\Omega_\Lambda=\frac{\rho_\Lambda}{\rho_{cr}}=\frac{\Lambda
c^2}{3H_0^2},\ \ \ \rho_{cr}=\frac{3H_0^2}{8\pi G_N}.
\end{equation}
can be calculated (here: $G_N$ and $h$  are the gravitation and
Planck constants). If Hubble constant $H_0$ = 70.5
($\text{kmsec}^{-1}$/ Mpc) \cite{26} then $\Omega_\Lambda\sim 0.73$.
An experimental value for $\Omega_\Lambda\sim 0.726 \pm 0.015$ was
recently obtained by the WMAP collaboration \cite{26}. We did
similar calculations for different $H_0$ in the article \cite{34} 10
years ago. For energy $\sim 150$ MeV (the end of the last phase
transition) the vacuum energy stopped to drop quickly and in further
the vacuum energy dropped very slowly. However, even at this moment
the large quantitative difference in densities of vacuum energy
between `hardness' and the modern value took place:
\begin{equation}\label{14}
(0.15 / 1.8\times10 ^{-12} )^ 4 \sim 5 \times 10 ^{43},\ \ \ \rho_{
DE} \sim (1.8 \times10 ^{-12} \text{GeV})^ 4 .
\end{equation}
This difference is very large but it is essentially smaller than 123
orders. The question is: how and why vacuum energy relaxed to the
modern value? Therefore, it is necessary to search another way for
understanding of this and it may be the holographic one.  Note that
at the moment of quark-hadron phase transition ($\sim$150 MeV) the
relation of components of the Universe was also hardened.

\section{Holographic principle}

The holographic theory of C. Balazs and I. Szapidi \cite{35} applied
to cosmology gives the following formula for the vacuum energy
density of the Universe in the holographic limit:
\begin{equation}\label{15}
\rho\leq3M_{Pl}^2/8\pi R^2.
\end{equation}
The vacuum energy density of the Universe is bounded by the inverse
area of its horizon. Here, important consequences of the holography
take place: energy is decreased by the linear size of the Universe;
energy density is decreased by its area. The authors of the article
\cite{35} used the Fischler- Susskind cosmic holographic conjecture
\cite{36} for which the entropy of the Universe ($S$) is limited by
its ``surface area'' measured in the Planckian units:
\begin{equation}\label{16}
S\leq\pi R^2M_{Pl}^2.
\end{equation}
In this case the connection between the vacuum energy density and a
number of quantum states of the Universe is arisen and then the
vacuum energy density following from equations (\ref{15}) and
(\ref{16}) is:
\begin{equation}\label{17}
\rho\leq3M_{Pl}^4/8S.
\end{equation}
Substituting the size of the observable Universe $R \sim 10^{ 28}$
cm in the formula (\ref{17}), we can get the vacuum energy density
of our Universe for $z =0$ (that is now) in the holographic limit
for $M_{Pl} =1$:
\begin{equation}\label{18}
\rho\sim4\times10^{-57}(GeV)^4.
\end{equation}
In other words, for expansion the vacuum energy is spent on
producing new quantum states. This value is significantly different
(10 orders) from the observable value of the vacuum energy density
$\rho\sim10^{ -47}(GeV) ^4$  but it is another side of the question.
Here it is necessary to give some explanation. General relativity is
the prime example of a holographic theory \cite{37}. But quantum
field theories, in the present form, are not holographic ones
\cite{35}. Therefore, in the quantum regime of the Universe
evolution the holographic concept does not work. The Universe came
in the classical (Friedmann) regime, probably, when $t \sim 10^{
-6}$ sec (corresponding to $E \sim 150$ MeV). $R_\text{QCD}$ was
then the causal horizon. If
\begin{equation}\label{19}
R_\text{QCD}\sim 3\times10^{4}cm,\ \ \
(R/R_\text{QCD})^2\sim10^{47}.
\end{equation}
Note that the holographic idea was first proposed in articles
\cite{37,38,39} and Ya. Bekenstein was the first who discussed this
idea applying them to black holes (BH) considering BH entropy (a
number of microstates) as a measure of information hidden in BH
\cite{40}. But the existence of the Universe horizon gives a
``strong argument'' supporting this holographic approach to the
solution of the cosmological constant problem. Here the increase of
entropy of the Universe (new quantum states) is evident. Besides,
both of these sizes ($10^{28}$ cm and $3\times10^4$ cm) are causal
horizons in the holographic thermodynamics in which a connection
between gravitation and thermodynamics takes place. Einstein's
equations are derived from the proportionality of entropy and the
horizon area together with the fundamental Clausius relation $dS =
dQ / T$  in which $dS$ is one quarter of the horizon area, $dQ$ and
$T$ are the energy flux across the horizon and Unruh temperature
seen by an accelerating observer inside the horizon \cite{37}. It is
non-equilibrium thermodynamics of space-time in some sense and here
thermodynamic derivation of the Einstein's equations appears. Even
more interesting moment is the statement that gravitation on a
macroscopic scale is a manifestation of thermodynamics of the
vacuum. It was the nontrivial idea of T. Jacobson \cite{37},
although S. Hawking \cite{39} many years ago underlined the
thermodynamic property of the de Sitter Universe to be similar to a
BH which written in the static coordinates.

The curious table can be made using cosmological parameters of the
seven-year WMAP data \cite{41} and the cosmological calculator of N.
Wright \cite{42} if $\Omega_\Lambda =0.73$; $\Omega_ m =0.27$; $H =
70.5$ ($\text{kmsec}^{-1}$/Mpc). Then, the density of the vacuum
energy in the classical regime as a function of redshift is:\\

\begin{widetext}
\begin{tabular}{|l|l|l|l|l|l|l|l|l|l|l|l|l|l|l|l|l|l|l|l|l|l|}\hline
$t=$&13.76&13.62&13.36&13.09&12.47&11.88&11.34&10.35&9.48&8.71&5.98&3.36&2.21
&1.58&1.2&0.49&0.18&0.1&47.9\\ \hline
$z=$&0&0.01&0.03&0.05&0.1&0.15&0.2&0.3&0.4&0.5&1&2&3&4&5&10&20&30&50
\\ \hline $10^{47}\rho=$&0.24&0.25&0.26&0.27&0.3&0.33&0.36&0.43&0.51&0.61&1.29&
4.12&9.62&19.13&31&197&1465&4687&21307\\
\hline
\end{tabular}
\end{widetext}
\begin{widetext}
\begin{tabular}{|l|l|l|l|l|l|l|l|l|l|l|l|l|l|l|l|l|l|l|l|l|l|}\hline
$t=$&16.8&1.37&0.44&0.22&0.13&25.4&6.9&1.8&0.3&75\\
\hline
$z=$&100&500&1000&1500&2000&5000&$10^4$&$2\times10^{4}$&$5\times10^{4}$&$10^5$\\
\hline
$10^{47}\rho=$&166666&$10^{40}\rho=2.5$&24.3&95.6&262&7212&97402&1431298&
$10^{32}\rho=0.51$&8.22\\
\hline
\end{tabular}
\end{widetext}
\begin{widetext}
\begin{tabular}{|l|l|l|l|l|l|l|l|l|l|l|l|l|l|l|l|l|l|l|l|l|l|}\hline
$t=$&0.76&$239\times10^3$&2396&25&0.27&0.003\\
\hline
$z=$&$10^6$&$10^7$&$10^8$&$10^9$&$10^{10}$&$10^{11}$\\
\hline $10^{32}\rho=$
&80128&$10^{24}\rho=7.29$&$10^{20}\rho=7.26$&$10^{16}\rho=6.67$&$10^{12}\rho=5.71$
 &$10^{8}\rho=4.62$  \\
\hline
\end{tabular}
\end{widetext}

where the time is in billion years from the creation of the Universe
up to $z=30$; from $z$=50 up to $z$=2000 the time is in million
years; from $z$=5000 up to $z$=50000 the time is in thousands years;
from $z=10^5$ up  to $z=10^6$ the time is in years; from $z=10^7$ up
to $z=10^{11}$ the time is in seconds. For calculation of density of
the vacuum energy, the simple approximate formulae have been used:
\begin{eqnarray}\label{20}
 \rho(z)&=& (3/8) M_{pl}^4 [R_\text{QCD}  /R(z)]^2\nonumber\\& =&0.375[(10^9/10^{56} )/
 r^2(z)]\nonumber\\
 &=&0.375 \times 10^{-47}/r^2(z) \ \ (GeV)^4.
\end{eqnarray}
For example, how can one get the density of the vacuum energy at
$z=0.5?$ For that one uses the cosmological calculator for
$\Omega_\Lambda=$0.73; $\Omega_ m$=0.27; $H_0$=70.5; $z$=0.5 and the
flat model \cite{42}. Then, the age at red shift $z$=0.5 was
$8.71\times10^9$ years (or $2.61\times10^{17}$sec). The causal
horizon was $R = 0.78 \times10^{ 28}$ cm and $r^2$(0.5) = 0.61.
Therefore, we have $\rho=0.375\times10^{-47}/0.61 \sim
0.61\times10^{-47}$. Note that during the time span from $z=3$
($t_3=2.21\times10^9$years) till $z$=0 ($t_0=13.76\times10^9$
years), the density of the vacuum energy  decreased 40 times, while
during the first $10^{-6}$ sec the Universe lost 78 orders owing to
the phase transitions. An initial part of this table may be checked
by the Ia supernova team in the following years \cite{43}.

\section{Conclusion}
There are the following probable points.
\begin{enumerate}
\item The relative content of the Universe components $\Omega_\Lambda$,
 $\Omega_\text{DM}$ and $\Omega_b$ had hardened in the
             first instants of the Universe evolution. The subsequent evolution led to decreasing absolute
             values of the component only.
\item    The cosmological constant relates the properties of microscopic
physics of a vacuum with
             the large scale physics.
\item    The vacuum energy density of our Universe was probably ($\sim M_{Pl}^4$) at the moment
 of its creation (it might be a fluctuation in the high symmetrical quantum vacuum of a
 multiverse)
 \item   Supersymmetry is broken if and only if the cosmological constant is positive.
\item    In the first parts of first second of our Universe evolution,
there was a period of vacuum
       evolution when condensates of quantum fields carried negative contributions in the positive
       energy density of the vacuum. It was the period of the non-equilibrium vacuum in its quantum
       regime.  The 78 orders of the vacuum energy density from the 123 orders
were compensated before its
      `hardness'.
\item   The vacuum energy of the Universe `has hardened' for T $\sim$ 150 MeV
( the quark - hadron
        phase transition  started at temperature T $\sim$ 265 MeV).
        \item    Assuming that during the first parts of first second, the vacuum energy had lost 78 orders
             then in the next $4\times10^{ 17}$ sec  it has lost only 45 orders by the creation of new quantum states
             (that is the rate of loss of the vacuum energy has decreased  $10^{ 55}$  times).
      \item  Of course, traces of relativistic phase transitions are not present nowadays although fractality
             in the distribution of the baryon component might be produced only phase transitions [44].
\item    The problem of the cosmological constant is probably solved by the implementation of
       the holographic principle to the `equilibrium vacuum' after its practical `hardness'.
\item     A holographic idea extended to all past history of our Universe's evolution from $z
=\infty$ to $z = 0$ was already considered in the article \cite{35}.
But it is not probably that the holographic
 principle may be applied to very early stages of the Universe evolution since an inflation
 phase was in that moment. Of course, the quantum regime of evolution took place in any case.
 \item   AdS/CFT correspondence, which states that all information about a gravitational system in
  any space region is encoded in its boundary, provides the strongest support to the
  holographic principle. This was noted by J. Maldacena 12 years ago \cite{45}.
  \item    Probably, Bekenstein's thermodynamics of BH may be a trace of the ``thermal nature'' of the
  Minkowski vacuum.
 \item    Introduced by E. Verlinde an entropic force \cite{46} as the specific microscopic force of
              space-time is a very natural  physical point of view. Here, classical gravity results from a
 thermodynamic approach.
\item    We could not get exactly 45 orders  (we have got 47 orders)
but this moment is not critical to the cosmological constant
problem. We are only showing in this publication that the crisis of
astrophysics connected with the cosmological constant is absent.
\end{enumerate}

Of course, some unsolved problems remain. We do not know well even
the equation of state of the dark energy which gradually losses its
dark status in favor of the vacuum energy (now $1+w = 0.013_{
-0.068}^{ +0.066}$ (0.11 syst)) [47]. The evidence for cosmic
acceleration exists now at the very high level  (more than 50
$\sigma$ [48])) but no evidence for DE evolution from a global
analysis of cosmological data [49]. Therefore, a scalar field must
be probably included for the best coincidence with cosmological data
although it will be a more complicate physical situation. If in this
case DE is given by a dynamical scalar field then it may have a
direct interaction with other material fields of the Universe, in
particular with cold dark matter [50]. Practically everything about
the dark energy including DE projects can be found in the last
detailed review [51] and in the article [43].

Finally, note that other approaches to dark energy modeling, which
predict $w \neq -1$ and f(R) gravity as well as proposals for
control experiments are intensively investigated [52-60]. Lastly it
is important to mention recent articles discussing the holographic
principle in cosmology [61-64]. Also G. Vereshkov recently noted the
important fact that the cosmological constant may be by Sakharov's
inducing gravitation [65].

\subsubsection*{Acknowledgment}
I am pleasure to thank Mubasher Jamil who has helped me very much
for preparation of this article to publication. Also it is pleasure
to thank S. Khakshournia who drew attention to the article\cite{66}.

\end{document}